\documentclass{PoS}

\def\als{\alpha_s} 
\def\MS{\overline{\rm MS}}

\title{Determination of the strong coupling $\alpha_s$ from the QCD static energy}

\ShortTitle{Determination of the strong coupling $\alpha_s$ from the QCD static energy}

\author{\speaker{Xavier Garcia i Tormo}%
\\
Albert Einstein Center for Fundamental Physics. Institut f\"ur Theoretische Physik. Universit\"at Bern,
  Sidlerstrasse 5, CH-3012 Bern, Switzerland\\
        E-mail: \email{garcia@itp.unibe.ch}}

\abstract{We obtain a determination of the strong coupling $\alpha_s$ in quantum
chromodynamics, by comparing perturbative calculations for the
short-distance part of the static energy with lattice
computations. Our result reads $\alpha_s\left(1.5{\rm
    GeV}\right)=0.326\pm0.019$, and when evolved to the scale $M_Z$
(the $Z$-boson mass) it corresponds to $\alpha_s\left(M_Z\right)=0.1156^{+0.0021}_{-0.0022}$.}

\FullConference{Xth Quark Confinement and the Hadron Spectrum\\
                 8-12 October 2012\\
                 TUM Campus Garching, Munich, Germany}

\begin{document}
This talk is based on Ref.~\cite{Bazavov:2012ka}. The reader is
referred to that paper for additional explanations.

\section{Introduction}
The energy between a static (i.e. infinitely heavy) quark and a static
antiquark separated a distance $r$ is known as the quantum
chromodynamics (QCD) static energy, $E_0(r)$, and is a basic object
to understand the dynamics of the theory. As it is well known, one
can distinguish a long-distance part and a short-distance part of the
static energy. The long-distance part encodes the confining dynamics
of the theory, whereas the short-distance part can be computed in
perturbation theory. On the other hand, one can use lattice QCD
simulations to compute the static energy in both (short- and long-
distance) regimes. Here we want to focus only on the short-distance
part of the static energy, where the perturbative weak-coupling
approach is expected to be reliable. In particular, we want to compare
state-of-the-art perturbative calculations with the lattice
results. This comparison allows us to obtain a determination of the
strong coupling $\alpha_s$, which is the main outcome of the present analysis.

\section{Perturbative calculation}
At short distances, $E_0(r)$ is given at leading order by the Coulomb
potential (with the adequate color factor), $E_0(r)\sim
-C_F\alpha_s/r$ (where $C_F=(N_c^2-1)/(2N_c)$, and $N_c$ is the number
of colors), and then we have corrections to this result. At
present the static energy is known including terms up to order $\als^{4+n}\ln^n\als$ with $n\ge
0$
\cite{Brambilla:2010pp,Pineda:2011db,Smirnov:2009fh,Anzai:2009tm,Brambilla:2009bi,Smirnov:2008pn,Brambilla:2006wp}
(a level of accuracy which we refer to as next-to-next-to-next-to
leading logarithmic -N$^3$LL-). The presence of $\ln\als$ terms in the
expansion of the static energy is due to the virtual emissions of gluons with energy of order $E_0$ (the so-called ultrasoft gluons), that can change the color state of the
quark-antiquark pair from singlet to octet and vice versa
\cite{Appelquist:1977es,Brambilla:1999qa}. Detailed expressions for
the static energy at this level of accuracy were given in Ref.~\cite{Brambilla:2010pp} (and
references therein; see also \cite{Tormo:2012rc} for explicit numerical expressions), and we will not reproduce them here.

\section{Lattice computation}
The static energy has been recently calculated in
$2+1$ flavor lattice QCD \cite{Bazavov:2011nk}. This computation used
a combination of tree-level improved gauge action and highly-improved
staggered quark action \cite{Follana:2006rc}; it employed the physical
value for the strange-quark mass $m_s$ and light-quark masses equal to
$m_s/20$ (corresponding to pion masses of about 160MeV). It was
performed for a wide range of gauge couplings, $5.9\le \beta \equiv
10/g^2 \le 7.28$. At each value of the gauge coupling one calculates
the scale parameters $r_0$ and $r_1$, defined in terms of the static energy $E_0(r)$ as follows \cite{Sommer,milc04}
\begin{equation}
r^2 \frac{d E_0(r)}{d r}|_{r=r_0}=1.65,~~~r^2 \frac{d E_0(r)}{d r}|_{r=r_1}=1.
\end{equation}
The values of $r_0$ and $r_1$ were given in Ref. \cite{Bazavov:2011nk} for each $\beta$.
The above range of gauge couplings corresponds to lattice spacings 
$1.909/r_0 \le a^{-1} \le 6.991/r_0$. Using the most recent value $r_0=0.468\pm0.004$ fm
\cite{Bazavov:2011nk} we get $0.805\,{\rm GeV}<a^{-1}< 2.947\,{\rm GeV}$. 
Thus we can study the static energy down to distances $r=0.14r_0$ or
$r\simeq 0.065$ fm. 
For the comparison with perturbation theory the most relevant data set
is the one that corresponds to $\beta = 6.664, 6.740, 6.800, 6.880, 6.950, 7.030 , 7.150,
7.280$, which is what we use here. The static energy can be calculated
in units of $r_0$ or $r_1$. Since the static energy has an additive
ultraviolet renormalization (self energy of the static sources) one needs to normalize the results calculated at different lattice spacings to a common value at a
certain distance (or alternatively one can take a derivative and
compute the force). The static energy is fixed, in units of $r_0$, to 0.954 at
$r = r_0$ \cite{Bazavov:2011nk}. At distances comparable to the
lattice spacing the static energy suffers from lattice artifacts. To
correct for these artifacts we use tree level improvement. That is, from the lattice Coulomb potential 
\begin{equation}
C_L(r)=\int\frac{d^3 k}{(2 \pi)^3} D_{00}(k_0=0,\vec{k}) e^{i \vec{k} \vec{r}},
\end{equation}
we can define the improved distance $r_I=(4 \pi C_L(r))^{-1}$ for each
separation $r$. Here $D_{00}$ is the tree level gluon propagator for the
$a^2$ improved gauge action. The tree level improvement amounts to replacing
$r$ by $r_I$ \cite{Necco:2001xg}. Alternatively following Ref. \cite{milc04,ukqcd}
we fit the lattice data at short distances to
the form $const-a/r +\sigma r+a' (1/r-1/r_I)$ and subtract the last term from
the lattice data. Since the data at the shortest distances that we use
(for each $\beta$)
correspond to a separation of one lattice spacing, it is important to
check that the way we are using to correct lattice artifacts is
working properly. In that sense, we have found that both methods of correcting for lattice
artifacts lead to the same results within errors of the calculations. Furthermore,
the static energies calculated for different lattice spacings agree well with
each other after the removal of lattice artifacts, and when one puts
all the data together it seems to lie on a single curve, even at short
distances, indicating that the above procedure of removing the lattice artifacts works.

\section{Comparing lattice and perturbation theory: $\alpha_s$ extraction}
We can now compare the lattice results with the perturbative
expressions, and use the comparison to extract the value of the QCD
scale $\Lambda_{\MS}$ (in the $\MS$ scheme). In order to obtain this
extraction, we assume that
perturbation theory (after implementing a cancellation of the leading
renormalon singularity) is enough to describe lattice data in the range of
distances we are considering (we use lattice data for $r<0.5r_0$, and
since we have lattice data points down to $r=0.14r_0$, this means that we
are studying the static energy in the 0.065~fm$\lesssim r \lesssim$0.234~fm distance range, in physical units). Then we search for the values of
$\Lambda_{\MS}$ that are allowed by lattice data; the guiding
principle to do that is that the agreement with lattice should improve when the
perturbative order of the calculation is increased.

As it was already mentioned above, in the perturbative calculation of
the energy one needs to implement a scheme that cancels the leading
renormalon singularity \cite{Beneke:1998rk,Hoang:1998nz}. This kind of schemes
introduce an additional dimensional scale in the problem (that we
denote as $\rho$). We implement the renormalon cancellation according
to the RS-scheme described in Ref.~\cite{Pineda:2001zq}. The static
energy in this scheme is given by
\begin{equation}
E_0^{\rm RS}(r,\rho)=E_0^{\MS}(r)-\textrm{RS\small{subtr.}}(\rho),
\end{equation}
where the subtraction term on the right-hand side cancels the leading
renormalon singularity of $E_0^{\MS}(r)$; the explicit expression for
$\textrm{RS\small{subtr.}}(\rho)$ is given, for instance, in Eq.~(7) of
Ref.~\cite{Tormo:2012rc}. The scale $\rho$
has, in this case, a natural value which corresponds to the center of
the range where we compare with lattice data (i.e. around 1.5~GeV), but any value around
that one cancels the renormalon and is, therefore, allowed.

\subsection{Central value for $r_0\Lambda_{\MS}$}
To obtain our central value for $r_0\Lambda_{\MS}$ we use the
following procedure:
\begin{enumerate}
\item We let $\rho$ vary by $\pm 25\%$ around its natural value.
\item For each value of $\rho$ and at each order in the perturbative
  expansion of the static energy, we perform a fit to the lattice
  data ($r_0\Lambda_{\MS}$ is the parameter of each of the fits).
\item We select those $\rho$ values for which the reduced $\chi^2$ of
  the fits decreases when increasing the number of loops of the perturbative calculation.
\end{enumerate}
Then we consider the set of $r_0\Lambda_{\MS}$ values in the $\rho$
range we have obtained and take their average, using the inverse
reduced $\chi^2$ of each fit as weight. From that, we obtain our
central value for $r_0\Lambda_{\MS}$. The value we obtain at 3 loop
with leading resummation of the ultrasoft logarithms is
$r_0\Lambda_{\MS}=0.70$, which will be our final number for the
central value. The perturbative expressions at N$^3$LL accuracy (i.e. 3 loop
with sub-leading resummation of the ultrasoft logarithms) are also
known (as mentioned before) but in this case an additional constant
appears in the expressions (due to the structure of the
renormalization group equations at this order \cite{Brambilla:2009bi}). This additional constant would also need
to be fitted to the lattice data (i.e. one has a two-parameter fit in
this case). When we do that we find that the
$\chi^2$ as a function $r_0\Lambda_{\MS}$ is very flat, and we cannot
improve the extraction by including these higher order terms. In
principle, more precise lattice data, and/or data at shorter distances
might allow for an improvement in that respect. 

\subsection{Error estimate}
Having obtained our
central value for $r_0\Lambda_{\MS}$ we now need to assign an error to
it. We want the error to reflect the uncertainties associated to the
neglected higher-order terms in the perturbative expansion of $E_0(r)$. To achieve
that, we consider: (i) the weighted standard
deviation in the set of $\rho$ values we found above, and (ii) the
difference with the weighted average computed at the previous
perturbative order. (Note that, starting at
two-loop order, one can decide whether one wants to perform the
resummation of the ultrasoft logarithms or not. To assign the error we
take whichever difference -with or without resummation in the previous
order- is larger. This amounts to not making any
assumption about the necessity or not to resum these logarithms). We then add
the two errors linearly (term (ii) turns out to be the dominant one). 

Additionally, we also redo the
analysis with alternative weight assignments ($p$-value, and constant
weights); we obtain compatible results. In the final result, we quote and error that
covers the whole range spanned by the three analyses. As an additional
cross-check, we can compare the analysis performed with the static energy normalized in
units of $r_0$ (our default choice) and the one with the static energy
normalized in units of $r_1$. We find that the two analyses give
consistent results. (Note that the values for the static energy in both
cases -i.e. in units of $r_0$ or $r_1$- come from the same lattice data set
in terms of $r/a$; but the error analysis in the normalization of the
energy for each lattice spacing is different in the two
cases. Therefore, one cannot obtain $E_0(r/r_1)$ from $E_0(r/r_0)$ by a
trivial rescaling, and it is in this sense that the analysis with the
scale $r_1$ provides a cross-check of the result).

\subsection{Final result for $\alpha_s$}
Our final result reads
\begin{equation}\label{eq:Lambda}
r_0\Lambda_{\MS}=0.70\pm0.07,
\end{equation}
which corresponds to
\begin{equation}\label{eq:as}
\alpha_s\left(\rho=1.5{\rm GeV},n_f=3\right)=0.326\pm0.019\quad\rightarrow\quad \alpha_s\left(M_Z,n_f=5\right)=0.1156^{+0.0021}_{-0.0022},
\end{equation}
where we used the $r_0$ value from Ref.~\cite{Bazavov:2011nk} to
obtain $\alpha_s(1.5{\rm GeV})$ from Eq.~(\ref{eq:Lambda}), and then
evolved it to the $Z$-mass scale, $M_Z$, using the \verb|Mathematica| package \verb|RunDec| \cite{Chetyrkin:2000yt} (4 loop running, with the charm quark mass equal to
1.6 GeV and the bottom quark mass equal to 4.7 GeV).

\subsection{Comparison with other recent $\alpha_s$ determinations}
\begin{figure}
\centering
\includegraphics[width=14cm]{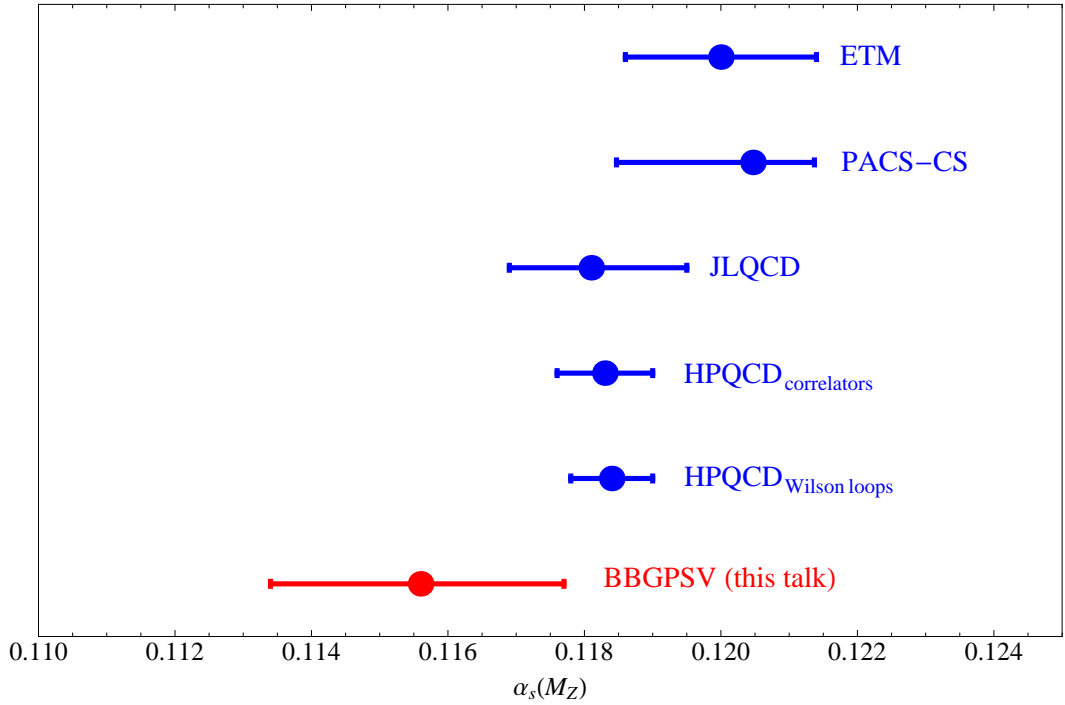}
\caption{Comparison of our result for $\alpha_s(M_Z)$ with other recent lattice determinations. The
  references are: HPQCD \cite{McNeile:2010ji}, JLQCD
  \cite{Shintani:2010ph}, PACS-CS \cite{Aoki:2009tf}, ETM \cite{Blossier:2012ef}.
}\label{fig:compalslatt}
\end{figure}

There are several recent determinations of $\alpha_s$ that also
employ comparisons with lattice data. These include analyses that use:
observables related to Wilson loops (but not the static energy)
\cite{Davies:2008sw,McNeile:2010ji}, moments of heavy quark
correlators \cite{Allison:2008xk,McNeile:2010ji}, the vacuum
polarization function \cite{Shintani:2010ph}, the Schr\"odinger
functional scheme \cite{Aoki:2009tf}, and the ghost-gluon coupling
\cite{Blossier:2012ef}. They deliver numbers that are mostly
compatible with our result, although our central value is a bit lower
than those of the other lattice determinations (see
Fig.~\ref{fig:compalslatt} for a graphical comparison).

\begin{figure}
\centering
\includegraphics[width=14cm]{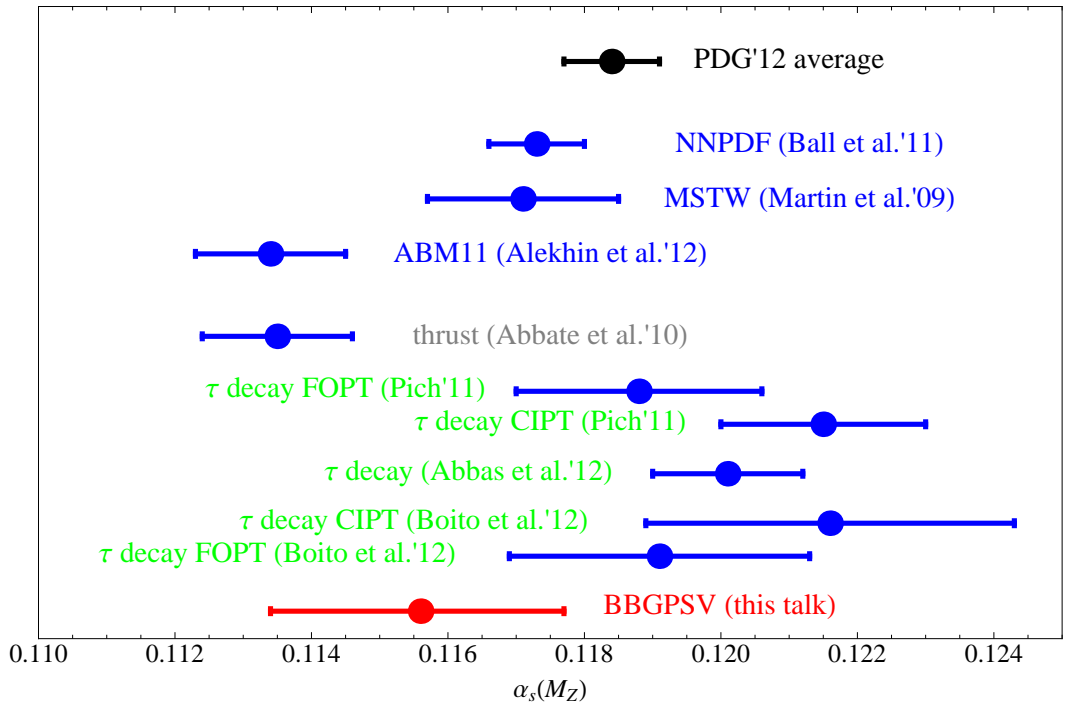}
\caption{Comparison of our result for $\alpha_s(M_Z)$ with a few other recent $\alpha_s$
  determinations. We include results from $\tau$ decays (Boito {\it et
    al.}~\cite{Boito:2012cr}; Abbas {\it et al.}~\cite{Abbas:2012py};
  Pich~\cite{Bethke:2011tr}), thrust (Abbate {\it et
    al.}~\cite{Abbate:2010xh}), and PDF fits (ABM11~\cite{Alekhin:2012ig},
  MSTW~\cite{Martin:2009bu}, NNPDF~\cite{Ball:2011us}), along with the
  PDG average \cite{Beringer:1900zz} for reference.
}\label{fig:compals}
\end{figure}

In Fig.~\ref{fig:compals} we compare our result for $\alpha_s(M_Z)$ in Eq.~(\ref{eq:as})
with a few other recent $\alpha_s$ determinations that use other
techniques (i.e. non-lattice determinations);
we include results coming from $\tau$ decays, thrust,
and parton distribution function (PDF) fits\footnote{Note that the
  errors in the results from PDF fits do not include effects from
  the unknown higher-order perturbative corrections. This theoretical
  uncertainty is difficult to assess and has not been addressed in
  detail so far. It is expected to be roughly of the same order as the quoted errors.}, along with the Particle
Data Group (PDG) average (the comparison is
not exhaustive, we just show a few other recent results in the
figure, meant to illustrate where our result lay with respect to
recent $\alpha_s$ extractions).

It is also worth remarking that our $\alpha_s$ determination is
performed at a scale of around 1.5~GeV, and therefore constitutes an
important new ingredient to further test the running of the strong
coupling (see Fig.~2 of Ref.~\cite{Bazavov:2012ka} for a graphical
comparison of different determinations of $\alpha_s$ as a function of
the energy scale $Q$ where they are performed).

\section{Conclusions}
To summarize, in this work we have compared perturbative calculations
for the QCD static
energy at short distances with lattice computations. We find that
perturbation theory (after canceling the leading renormalon
singularity) is able to describe the short-distance part of the static
energy computed in $2+1$ flavor lattice QCD (see Fig.~1 of
Ref.~\cite{Bazavov:2012ka} for a comparison of the different orders of
accuracy of the perturbative result and the lattice data). We
exploited this fact to obtain a determination of the strong coupling
$\alpha_s$. Our extraction is at three-loop accuracy
(including resummation of the leading ultrasoft logarithms) and is
performed at a scale of 1.5~GeV. When we evolve the result to the
scale $M_Z$ it corresponds to $\alpha_s\left(M_Z\right)=0.1156^{+0.0021}_{-0.0022}$.

\acknowledgments
It is a pleasure to thank Alexei Bazavov, Nora Brambilla, P\'eter
Petreczky, Joan Soto, and Antonio Vairo for collaboration on the work
reported in this talk.

\end{document}